\documentclass[aps,prb,onecolumn,showpacs]{revtex4}
\usepackage{latexsym}
\usepackage{amssymb}
\usepackage{graphicx}
\begin{document}

\title{Quantum transport through double-dot Aharonov-Bohm interferometry in Coulomb blockade regime}
\author{Jing Ma, Bing Dong, and X. L. Lei}
\address{Department of Physics, Shanghai Jiaotong University, 1954 Huashan Road, Shanghai 200030, China}
\date{\today}

\begin{abstract}
Transport through two quantum dots laterally embedded in Aharonov-Bohm
interferometry with infinite intradot and arbitrary interdot Coulomb repulsion
is analyzed in the weak coupling and Coulomb blockade regime.
By employing the modified quantum rate equations and the slave-boson approach,
we establish a general dc current formula at temperatures higher than the Kondo temperature
for the case that the spin degenerate levels of two dots are close to each other.
We examine two simple examples for identical dots - no doubly occupied states and no empty state.
In the former, completely destructive coherent transport and phase locking
appear at magnetic flux $\Phi=\Phi_{0}/2$ and $\Phi=0$ respectively; in the latter, partially coherent transport exhibits
an oscillation with magnetic flux having a period of $\Phi_0$.
\end{abstract}
\pacs{73.21.La Quantum dots, 73.23.-b Electronic transport in mesoscopic systems, 73.23.Hk Coulomb blockade and single-electron tunneling.}
\maketitle

\section{Introduction}
Quantum dot (QD), a tiny engineered device accommodating a single
electron or a few ones in three-dimensionally confined space, acts
not only as a crucial ingredient for the realization of solid state quantum computation
but also as a convenient tool to explore the effect of strong correlation manifested by discrete energy levels.
Very rich phenomena, such as resonant tunneling, Coulomb blockade and the Kondo effect, arise in different
circumstances to intrigue experimentists and theorists. To address the phase coherence of the transport, the QD or QDs are embedded in
various Aharonov-Bohm (AB) geometries. Till now single dot \cite{E1, E2, E3, E4, E5}
or double dots \cite{E6, E7, E8} in two-terminal AB interferometer have been
realized in experiments and the current oscillation of magnetic flux has been observed.
\begin{figure}[h]
\includegraphics [width=0.5\textwidth,clip] {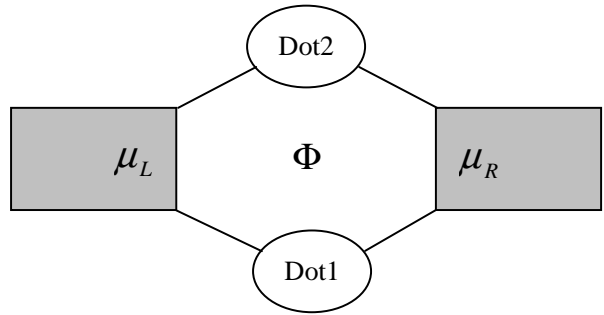}
\caption{Schematic of a double-dot system in a parallel configuration between two leads in the presence 
of Aharonov-Bohm magnetic flux.}
\label{fig1}
\end{figure}

In this paper, we consider two quantum dots parallely connected to two metallic reservoirs with
magnetic flux penetrating the enclosed area. Due to the complexity of this system,
most of the previous research concentrated on the noninteracting case
to give the exact and general results of transport and to find effects of magnetic flux, position and difference
of QDs energy levels, and band widths of coupling strength \cite{TVS, BKU, KKA, JKO, ZTJ, MLL, ZMB}.
There were only a few thoretical studies of the interacting systems, which focused
either on intradot correlation with spin for two identical dots \cite{HAK, JKO, DLO},
or on interdot correlation without spin for two different dots \cite{LEV, DBO},
or on the case having large intradot and interdot correlation without spin for two identical dots \cite{JKO}.
Ref.\cite{DBOE} is an exception where the intradot and interdot Coulomb repulsion, spin configuration and disparity
of two dot levels were discussed at zero temperature covering
the strong coupling Kondo regime. The present paper also deals with
intradot and interdot Coulomb correlation, spin configuration and level disparity
but is concerned with the Coulomb blockade regime at temperatures higher than the Kondo temperature.

For the description of quantum transport through double-dot system, the "classical" rate equations must be
modified for nondiagonal density matrix elements responsible for transitions between isolated quantum
states \cite{Nazarov, Korotkov, Goan, Elattari, Stoof}. Recently a modified quantum equations have been derived \cite{Dong} to study the transport of
an interacting system using the slave-boson technique introduced by Zou and Anderson \cite{Zou}
and incorporating the nonequilibrium Green's functions. The solutions are equivalent to the lowest-order gradient expansion, which is a good approximation
for sequential resonant tunneling \cite{Davies}, and essentially accordant with previous analyses.
Utilizing this kind of quantum rate equations, one is able to discuss the Coulomb correlation effect at arbitrary temperature.
In the present paper, we apply this method to analyze a double-dot AB interferometer with interdot and intradot Coulomb repulsion,
considering the lowest-order of the dot-lead coupling strength in transport.

\section{Formulation}
The Hamiltonian of two tunneling coupled quantum dots parallely connected to left and right leads
with the presence of magnetic flux is described by the genetic tunneling model:
\begin{eqnarray}
H&=&\sum\limits_{{\alpha}k\sigma}\epsilon_{{\alpha}k\sigma}c^{\dag}_{{\alpha}k\sigma}c_
{{\alpha}k\sigma}
+\epsilon_{1}\sum\limits_{\sigma}c^{\dag}_{1\sigma}c_{1\sigma}+Un_{1\uparrow}
n_{1\downarrow}
+\epsilon_{2}\sum\limits_{\sigma}c^{\dag}_{2\sigma}c_{2\sigma}+Un_{2\uparrow}
n_{2\downarrow}
+U'\sum\limits_{\sigma\sigma'}n_{1\sigma}n_{2\sigma'}\\\nonumber
&+&\sum\limits_{k\sigma}(t_{L1\sigma}c^{\dag}_{Lk\sigma}c_{1\sigma}+t_{R1\sigma}c^{\dag
}_{Rk\sigma}c_{1\sigma}+h.c.)
+\sum\limits_{k\sigma}(t_{L2\sigma}c^{\dag}_{Lk\sigma}c_{2\sigma}+t_{R2\sigma}c^{\dag}_
{Rk\sigma}c_{2\sigma}+h.c.) ,
\end{eqnarray}
where $c^{\dag}_{1(2)\sigma}$ ($c_{1(2)\sigma}$) and $c^{\dag}_{\alpha k\sigma}$ ($c_{\alpha k\sigma}$) are creation
(annihilation) operators of electrons in the dots $1(2)$ and in the left and right leads ($\alpha=L, R$) with spin $\sigma$.
Each dot has a single spin degenerate orbital level $\epsilon_{1(2)}$ and an infinite on-site Coulomb repulsion
$U$ and simultaneously there is an arbitrary finite interdot electrostatic correlation $U'$ between them.
We only consider the two dot levels are very close to each other,
$\epsilon_{2(1)}=\epsilon_{d}{\pm}\epsilon/2$, with a small variation $\epsilon$.
The effect of AB flux $\Phi$ is taken into account in the tunneling amplitude $t_{\alpha 1(2)\sigma}$
by $\varphi=2\pi{\Phi}/{\Phi_0}$ with the flux quantum $\Phi_0=h/e$. In the Peierls gauge, one
generally chooses $t^{*}_{L1\sigma}=t_{L2\sigma}=t^{*}_{R2\sigma}=t_{R1\sigma}=|t|e^{i\varphi/4}$ 
\cite{BKU, JKO}.
The $\alpha$th lead is supposed to be Fermi liquids in equilibrium state and has the same coupling strength function
with two dots $\Gamma_{\alpha\sigma}(\omega)=2\pi\sum_{k\alpha}|t_{\alpha 1(2)\sigma}|^2\delta(\omega-\epsilon_{\alpha k\sigma})$.

According to the slave-particle approach originated by Zou and Anderson,
we introduce auxiliary operators $e^{\dag},\  f^{\dag}_{1\sigma}\  (f^{\dag}_{2\sigma}),\  d^{\dag}_{\sigma\sigma'}$
to stand for the possible states of two dots as a whole: empty state $|0{\rangle}_{1}|0{\rangle}_{2}$,
singly occupied state $|\sigma{\rangle}_{1}|0{\rangle}_{2}\  (|0{\rangle}_{1}|\sigma{\rangle}_{2})$,
and doubly occupied state $|\sigma{\rangle}_{1}|\sigma'{\rangle}_{2}$ respectively.
In the slave-boson representation, the electron operators of each dot are substituted by the slave-boson operators $e^{\dag}$, $d^{\dag}$
and the pseudo-fermion operators $f^{\dag}_{1\sigma}$, $f^{\dag}_{2\sigma}$ \cite{Zou, Dong}:
\begin{eqnarray}
c_{1\sigma}=e^{\dag}f_{1\sigma}+\sum\limits_{\sigma'}f^{\dag}_{2\sigma'}
d_{\sigma\sigma'},\
c_{2\sigma}=e^{\dag}f_{2\sigma}+\sum\limits_{\sigma'}f^{\dag}_{1\sigma'}
d_{\sigma'\sigma},
\end{eqnarray}
with the completeness constraint
$e^{\dag}e+\sum_{\sigma}(f^{\dag}_{1\sigma}f_{1\sigma}+f^{\dag}_{2\sigma}
f_{2\sigma})+\sum_{\sigma\sigma'}d^{\dag}_{\sigma\sigma'}d_{\sigma\sigma'}=1$,
for which these operators must be correctly quantized to make the sum rule for the physical electron valid
and the commutators between them are satisfying \cite{Guillou, Dong}:
\begin{eqnarray}
ee^{\dag}&=&1,\ d_{{\sigma}_{1}{\sigma}_{2}}d^{\dag}_{{\sigma'}_{1}{\sigma'}_{2}}={\delta}_{{\sigma}_{1
}{\sigma'}_{1}}{\delta}_{{\sigma}_{2}{\sigma'}_{2}},
\ f_{i\sigma}f^{\dag}_{j\sigma'}={\delta}_{ij}{\delta}_{\sigma\sigma'},\\\nonumber
ed^{\dag}_{\sigma\sigma'}&=&ef^{\dag}_{i\sigma}=f_{i\sigma}e^{\dag}=f_{i\sigma}d^{\dag}_{\sigma'\sigma''}=d_{\sigma\sigma'}e^{\dag}=d_{\sigma'\sigma''}f^{\dag}_{i\sigma}=0.
\end{eqnarray}

So the effective Hamiltonian is written in terms of these instrumental state operators:
\begin{eqnarray}
H_{eff}&=&\sum\limits_{{\alpha}k\sigma}\epsilon_{{\alpha}k\sigma}c^{\dag}_{{\alpha}
k\sigma}c_{{\alpha}k\sigma}
+\epsilon_{1}\sum\limits_{\sigma}f^{\dag}_{1\sigma}f_{1\sigma}
+\epsilon_{2}\sum\limits_{\sigma}f^{\dag}_{2\sigma}f_{2\sigma}
+(\epsilon_{1}+\epsilon_{2}+U')\sum\limits_{\sigma\sigma'}d^{\dag}_{\sigma\sigma'}
d_{\sigma\sigma'}\\\nonumber
&+&\sum\limits_{k\sigma}[t_{L1\sigma}c^{\dag}_{Lk\sigma}(e^{\dag}f_{1\sigma}+
\sum\limits_{\sigma'}f^{\dag}_{2\sigma'}d_{\sigma\sigma'})
+t_{R1\sigma}c^{\dag}_{Rk\sigma}(e^{\dag}f_{1\sigma}+\sum\limits_{\sigma'}
f^{\dag}_{2\sigma'}d_{\sigma\sigma'})+h.c.]\\\nonumber
&+&\sum\limits_{k\sigma}[t_{L2\sigma}c^{\dag}_{Lk\sigma}(e^{\dag}f_{2\sigma}+
\sum\limits_{\sigma'}f^{\dag}_{1\sigma'}d_{\sigma'\sigma})
+t_{R2\sigma}c^{\dag}_{Rk\sigma}(e^{\dag}f_{2\sigma}+\sum\limits_{\sigma'}
f^{\dag}_{1\sigma'}d_{\sigma'\sigma})+h.c.],
\end{eqnarray}
and the density matrix elements are expressed as $\rho_{00}=|0{\rangle}_{1}|0{\rangle}_{22}{\langle}0|_{1}{\langle}0|=e^{\dag}e$, 
$\rho_{11\sigma}=|\sigma{\rangle}_{1}|0{\rangle}_{22}{\langle}0|_{1}{\langle}\sigma|
=f^{\dag}_{1\sigma}f_{1\sigma}$,
$\rho_{22\sigma}=|0{\rangle}_{1}|\sigma{\rangle}_{22}{\langle}\sigma|_{1}{\langle}0|
=f^{\dag}_{2\sigma}f_{2\sigma}$,
$\rho_{dd\sigma\sigma'}=|\sigma{\rangle}_{1}|\sigma'{\rangle}_{22}{\langle}
\sigma'|_{1}{\langle}\sigma|=d^{\dag}_{\sigma\sigma'}d_{\sigma\sigma'}$,
and $\rho_{12\sigma}=|0{\rangle}_{1}|\sigma{\rangle}_{22}{\langle}0|_{1}{\langle}\sigma|
=f^{\dag}_{2\sigma}f_{1\sigma}$,
$\rho_{21\sigma}=|\sigma{\rangle}_{1}|0{\rangle}_{22}{\langle}\sigma|_{1}{\langle}0|
=f^{\dag}_{1\sigma}f_{2\sigma}$.

Supposing the left and right leads are made from the identical material and
the effective coupling strength is constant for the energy range of interest
$\Gamma_{L,R\sigma}(\omega)=\Gamma_{L,R\sigma}$.
Starting from the equations of motion of the density matrix elements and after the Langreth analytic continuation to
decouple the dots and the leads interaction terms, the rate equations are acquired in the wide band limit:
\begin{eqnarray}
\dot{\rho}_{00}&=&\frac{-i}{2\pi}\int{d\omega}\sum\limits_{\sigma}[A_{\sigma}
(G^{>}_{e11\sigma}+G^{>}_{e22\sigma})+
C_{\sigma}(G^{<}_{e11\sigma}+G^{<}_{e22\sigma})+B_{\sigma}G^{>}_{e12\sigma}+
B^{*}_{\sigma}G^{>}_{e21\sigma}
+D_{\sigma}G^{<}_{e12\sigma}\\\nonumber&+&D^{*}_{\sigma}G^{<}_{e21\sigma}],
\end{eqnarray}
\begin{eqnarray}
\dot{\rho}_{11\sigma}&=&\frac{i}{2\pi}\int{d\omega}
(A_{\sigma}G^{>}_{e11\sigma}+{1/2}B_{\sigma}G^{>}_{e12\sigma}+{1/2}
B^{*}_{\sigma}G^{>}_{e21\sigma}
+C_{\sigma}G^{<}_{e11\sigma}+{1/2}D_{\sigma}G^{<}_{e12\sigma}+{1/2}
D^{*}_{\sigma}G^{<}_{e21\sigma})
\\\nonumber&-&\frac{i}{4\pi}\int{d\omega}\sum\limits_{\sigma'\sigma''}
[A_{\sigma'}(G^{>}_{11\sigma'\sigma\sigma''}+G^{>}_{11\sigma'\sigma''\sigma})
+B_{\sigma'}G^{''>}_{d21\sigma'\sigma''\sigma}+B^{*}_{\sigma'}
G^{>}_{d12\sigma\sigma'\sigma''}
+C_{\sigma'}(G^{<}_{11\sigma'\sigma\sigma''}
\\\nonumber&+&G^{<}_{11\sigma'\sigma''\sigma})
+D_{\sigma'}G^{''<}_{d21\sigma'\sigma''\sigma}+D^{*}_{\sigma'}
G^{<}_{d12\sigma\sigma'\sigma''}],
\end{eqnarray}
\begin{eqnarray}
\dot{\rho}_{22\sigma}&=&\frac{i}{2\pi}\int{d\omega}
(A_{\sigma}G^{>}_{e22\sigma}+{1/2}B_{\sigma}G^{>}_{e12\sigma}+{1/2}
B^{*}_{\sigma}G^{>}_{e21\sigma}
+C_{\sigma}G^{<}_{e22\sigma}+{1/2}D_{\sigma}G^{<}_{e12\sigma}+{1/2}
D^{*}_{\sigma}G^{<}_{e21\sigma})
\\\nonumber&-&\frac{i}{4\pi}\int{d\omega}\sum\limits_{\sigma'\sigma''}
[A_{\sigma'}(G^{>}_{22\sigma'\sigma\sigma''}+G^{>}_{22\sigma'\sigma''\sigma})
+B_{\sigma'}G^{''>}_{d21\sigma'\sigma\sigma''}+B^{*}_{\sigma'}
G^{>}_{d12\sigma''\sigma'\sigma}
+C_{\sigma'}(G^{<}_{22\sigma'\sigma\sigma''}
\\\nonumber&+&G^{<}_{22\sigma'\sigma''\sigma})
+D_{\sigma'}G^{''<}_{d21\sigma'\sigma\sigma''}+D^{*}_{\sigma'}
G^{<}_{d12\sigma''\sigma'\sigma}],
\end{eqnarray}
\begin{eqnarray}
\dot{\rho}_{12\sigma}&=&\frac{i}{4\pi}\int{d\omega}
[2A_{\sigma}G^{>}_{e12\sigma}+B^{*}_{\sigma}(G^{>}_{e11\sigma}+G^{>}_{e22\sigma})+
2C_{\sigma}G^{<}_{e12\sigma}+D^{*}_{\sigma}(G^{<}_{e11\sigma}+G^{<}_{e22\sigma})]
\\\nonumber&-&\frac{i}{4\pi}\int{d\omega}\sum\limits_{\sigma'\sigma''}
[A_{\sigma'}(G^{>}_{d21\sigma'\sigma''\sigma}+G^{'>}_{d21\sigma\sigma'\sigma''})
+B^{*}_{\sigma'}(G^{'>}_{d11\sigma'\sigma''\sigma}+G^{'>}_{d22\sigma\sigma'\sigma''})
+C_{\sigma'}(G^{<}_{d21\sigma'\sigma''\sigma}
\\\nonumber&+&G^{'<}_{d21\sigma\sigma'\sigma''})
+D^{*}_{\sigma'}(G^{'<}_{d11\sigma'\sigma''\sigma}+G^{'<}_{d22\sigma\sigma'\sigma''})]
+i(\epsilon_{2}-\epsilon_{1})\rho_{12\sigma},
\end{eqnarray}
\begin{eqnarray}
\dot{\rho}_{dd\sigma\sigma}&=&\frac{i}{4\pi}\int{d\omega}\sum\limits_{\sigma'}
[A_{\sigma}(G^{>}_{d11\sigma\sigma\sigma'}+G^{>}_{d11\sigma\sigma'\sigma}
+G^{>}_{d22\sigma\sigma\sigma'}+G^{>}_{d22\sigma\sigma'\sigma})
+B_{\sigma}(G^{''>}_{d21\sigma\sigma\sigma'}+G^{''>}_{d21\sigma\sigma'\sigma})
\\\nonumber
&+&B^{*}_{\sigma}(G^{>}_{d12\sigma'\sigma\sigma}+G^{>}_{d12\sigma\sigma\sigma'})
+C_{\sigma}(G^{<}_{d11\sigma\sigma\sigma'}+G^{<}_{d11\sigma\sigma'\sigma}
+G^{<}_{d22\sigma\sigma\sigma'}+G^{<}_{d22\sigma\sigma'\sigma})
+D_{\sigma}(G^{''<}_{d21\sigma\sigma\sigma'}
\\\nonumber&+&G^{''<}_{d21\sigma\sigma'\sigma})
+D^{*}_{\sigma}(G^{<}_{d12\sigma'\sigma\sigma}+G^{<}_{d12\sigma\sigma\sigma'})],
\end{eqnarray}
\begin{eqnarray}
\dot{\rho}_{dd\sigma\bar{\sigma}}&=&\frac{i}{4\pi}\int{d\omega}\sum\limits_{\sigma'}
[A_{\bar{\sigma}}(G^{>}_{d11\bar{\sigma}\sigma\sigma'}+
G^{>}_{d11\bar{\sigma}\sigma'\sigma})
+A_{\sigma}(G^{>}_{d22\sigma\bar{\sigma}\sigma'}+G^{>}_{d22\sigma\sigma'\bar{\sigma}})
+B_{\sigma}G^{''>}_{d21\sigma\bar{\sigma}\sigma'}\\\nonumber&+&B_{\bar{\sigma}}
G^{''>}_{d21\bar{\sigma}\sigma'\sigma}
+B^{*}_{\sigma}G^{>}_{d12\sigma'\sigma\bar{\sigma}}+B^{*}_{\bar{\sigma}}
G^{>}_{d12\sigma\bar{\sigma}\sigma'}
+C_{\bar{\sigma}}(G^{<}_{d11\bar{\sigma}\sigma\sigma'}+G^{<}_{d11\bar{\sigma}
\sigma'\sigma})
+C_{\sigma}(G^{<}_{d22\sigma\bar{\sigma}\sigma'}\\\nonumber&+&G^{<}_{d22\sigma\sigma'\bar{\sigma}})
+D_{\sigma}G^{''<}_{d21\sigma\bar{\sigma}\sigma'}+D_{\bar{\sigma}}
G^{''<}_{d21\bar{\sigma}\sigma'\sigma}
+D^{*}_{\sigma}G^{<}_{d12\sigma'\sigma\bar{\sigma}}+D^{*}_{\bar{\sigma}}
G^{<}_{d12\sigma\bar{\sigma}\sigma'}],
\end{eqnarray}
here $A_{\sigma}=f_L(\omega)\Gamma_{L\sigma}+f_R(\omega)\Gamma_{R\sigma}$,
$C_{\sigma}=[1-f_L(\omega)]\Gamma_{L\sigma}+[1-f_R(\omega)]\Gamma_{R\sigma}$,
$B_{\sigma}=f_L(\omega)\Gamma_{L\sigma}e^{-i\varphi/2}+f_R(\omega)
\Gamma_{R\sigma}e^{i\varphi/2}$,
$D_{\sigma}=[1-f_L(\omega)]\Gamma_{L\sigma}e^{-i\varphi/2}+[1-f_R(\omega)]
\Gamma_{R\sigma}e^{i\varphi/2}$,
with the Fermi distribution function $f_\alpha(\omega)=1/(e^{\beta(\omega-\mu_\alpha)}+1)$, $\mu_\alpha$
the chemical potential of the $\alpha$th lead at temperature $1/\beta$.

For a symmetric system $\Gamma_{L\sigma}=\Gamma_{R\sigma}=\Gamma_{\sigma}$, the dc current of the stationary state can be symmetrized $I=(I_L-I_R)/2$ and
$I_\alpha (\alpha=L,R)$ is obtained from the average of time derivative of electron number operator of the $\alpha$th lead:
\begin{eqnarray}
I&=&\frac{ie}{2h}\int{d\omega}\sum\limits_{\sigma}\Gamma_{\sigma}
\{[f_L(\omega)-f_R(\omega)][G^{>}_{e11\sigma}+G^{>}_{e22\sigma}
-G^{<}_{e11\sigma}-G^{<}_{e22\sigma}
+\sum\limits_{\sigma'\sigma''}
(G^{>}_{d11\sigma\sigma'\sigma''}+G^{>}_{d22\sigma\sigma'\sigma''}\\\nonumber
&-&G^{<}_{d11\sigma\sigma'\sigma''}-G^{<}_{d22\sigma\sigma'\sigma''})]
+\cos{\frac{\varphi}{2}}[f_L(\omega)-f_R(\omega)]
[G^{>}_{e12\sigma}+G^{>}_{e21\sigma}-G^{<}_{e12\sigma}-G^{<}_{e21\sigma}
+\sum\limits_{\sigma'\sigma''}
(G^{''>}_{d21\sigma\sigma'\sigma''}\\\nonumber&+&G^{>}_{d12\sigma''\sigma\sigma'}
-G^{''<}_{d21\sigma\sigma'\sigma''}-G^{<}_{d12\sigma''\sigma\sigma'})]
+i\sin{\frac{\varphi}{2}}[f_L(\omega)+f_R(\omega)]
[G^{>}_{e21\sigma}-G^{>}_{e12\sigma}+\sum\limits_{\sigma'\sigma''}
(G^{>}_{d12\sigma''\sigma\sigma'}\\\nonumber&-&G^{''>}_{d21\sigma\sigma'\sigma''})]
+2i\sin{\frac{\varphi}{2}}[1-\frac{f_L(\omega)+f_R(\omega)}{2}]
[G^{<}_{e21\sigma}-G^{<}_{e12\sigma}+\sum\limits_{\sigma'\sigma''}
(G^{<}_{d12\sigma''\sigma\sigma'}-G^{''<}_{d21\sigma\sigma'\sigma''})]\}.
\end{eqnarray}
This expression is similar with that of Ref.\cite{JKO} where the interdot correlation was not included.
The Green's functions quoted in the above are defined as follows:
\begin{eqnarray}
G_{eii\sigma}&=&<<e^{\dag}(t)f_{i\sigma}(t)|f^{\dag}_{i\sigma}(t')e(t')>>, 
G_{eij\sigma}=<<e^{\dag}(t)f_{i\sigma}(t)|f^{\dag}_{j\sigma}(t')e(t')>>,\\\nonumber
G_{d11\sigma'\sigma\sigma''}&=&<<f^{\dag}_{1\sigma}(t)d_{\sigma\sigma'}(t)
|d^{\dag}_{\sigma''\sigma'}(t')f_{1\sigma''}(t')>>, 
G^{'}_{d11\sigma'\sigma''\sigma}=<<f^{\dag}_{1\sigma''}(t)d_{\sigma''\sigma'}(t)
|d^{\dag}_{\sigma'\sigma}(t')f_{1\sigma}(t')>>,\\\nonumber
G_{d22\sigma'\sigma\sigma''}&=&<<f^{\dag}_{2\sigma}(t)d_{\sigma'\sigma}(t)
|d^{\dag}_{\sigma'\sigma''}(t')f_{2\sigma''}(t')>>, 
G^{'}_{d22\sigma\sigma'\sigma''}=<<f^{\dag}_{2\sigma}(t)d_{\sigma\sigma'}(t)
|d^{\dag}_{\sigma'\sigma''}(t')f_{2\sigma''}(t')>>,\\\nonumber
G_{d21\sigma'\sigma''\sigma}&=&<<f^{\dag}_{2\sigma''}(t)d_{\sigma'\sigma''}(t)
|d^{\dag}_{\sigma'\sigma}(t')f_{1\sigma}(t')>>, 
G^{'}_{d21\sigma\sigma'\sigma''}=<<f^{\dag}_{2\sigma}(t)d_{\sigma\sigma'}(t)
|d^{\dag}_{\sigma''\sigma'}(t')f_{1\sigma''}(t')>>,\\\nonumber
G^{''}_{d21\sigma'\sigma\sigma''}&=&<<f^{\dag}_{2\sigma}(t)d_{\sigma'\sigma}(t)
|d^{\dag}_{\sigma''\sigma'}(t')f_{1\sigma''}(t')>>, 
G_{d12\sigma\sigma'\sigma''}=<<f^{\dag}_{1\sigma}(t)d_{\sigma\sigma'}(t)
|d^{\dag}_{\sigma'\sigma''}(t')f_{2\sigma''}(t')>>.
\end{eqnarray}
In weak coupling approximation and with small level discrepancy $\epsilon$, the correlation Green's functions 
in the isolated two dot system are:
\begin{eqnarray}
G^{<0}_{eii\sigma}(\omega)&=&2{\pi}i\rho_{ii\sigma}\delta(\omega-\epsilon_d), 
G^{>0}_{eii\sigma}(\omega)=-2{\pi}i\rho_{00}\delta(\omega-\epsilon_d),\\\nonumber
G^{<0}_{eij\sigma}(\omega)&=&2{\pi}i\rho_{ij\sigma}\delta(\omega-\epsilon_d), 
G^{>0}_{eij\sigma}(\omega)=0,\\\nonumber
G^{<0}_{d11\sigma'\sigma\sigma''}(\omega)&=&\delta_{\sigma\sigma''}2{\pi}i
\rho_{dd\sigma\sigma'}\delta(\omega-\epsilon_d-U'), 
G^{>0}_{d11\sigma'\sigma\sigma''}(\omega)=-\delta_{\sigma\sigma''}2{\pi}i
\rho_{11\sigma}\delta(\omega-\epsilon_d-U'),\\\nonumber
G^{<0}_{d22\sigma'\sigma\sigma''}(\omega)&=&\delta_{\sigma\sigma''}2{\pi}i
\rho_{dd\sigma'\sigma}\delta(\omega-\epsilon_d-U'), 
G^{>0}_{d22\sigma'\sigma\sigma''}(\omega)=-\delta_{\sigma\sigma''}2{\pi}i
\rho_{22\sigma}\delta(\omega-\epsilon_d-U'),\\\nonumber
G^{'<0}_{d11\sigma'\sigma''\sigma}(\omega)&=&\delta_{\sigma\sigma'}
\delta_{\sigma\sigma''}2{\pi}i\rho_{dd\sigma\sigma}\delta(\omega-\epsilon_d-U'), 
G^{'>0}_{d11\sigma'\sigma''\sigma}(\omega)=-\delta_{\sigma\sigma'}
\delta_{\sigma\sigma''}2{\pi}i\rho_{11\sigma}\delta(\omega-\epsilon_d-U'),\\\nonumber
G^{'<0}_{d22\sigma\sigma'\sigma''}(\omega)&=&\delta_{\sigma\sigma'}
\delta_{\sigma\sigma''}2{\pi}i\rho_{dd\sigma\sigma}\delta(\omega-\epsilon_d-U'), 
G^{'>0}_{d22\sigma\sigma'\sigma''}(\omega)=-\delta_{\sigma\sigma'}
\delta_{\sigma\sigma''}2{\pi}i\rho_{22\sigma}\delta(\omega-\epsilon_d-U'),\\\nonumber
G^{<0}_{d21\sigma'\sigma''\sigma}(\omega)&=&0, 
G^{>0}_{d21\sigma'\sigma''\sigma}(\omega)=-\delta_{\sigma\sigma''}2{\pi}i
\rho_{12\sigma}\delta(\omega-\epsilon_d-U'),\\\nonumber
G^{'<0}_{d21\sigma\sigma'\sigma''}(\omega)&=&0, 
G^{'>0}_{d21\sigma\sigma'\sigma''}(\omega)=-\delta_{\sigma\sigma''}2{\pi}i
\rho_{12\sigma}\delta(\omega-\epsilon_d-U'),\\\nonumber
G^{''<0}_{d21\sigma'\sigma\sigma''}(\omega)&=&0, 
G^{''>0}_{d21\sigma'\sigma\sigma''}(\omega)=-\delta_{\sigma\sigma'}
\delta_{\sigma\sigma''}2{\pi}i\rho_{12\sigma}\delta(\omega-\epsilon_d-U'),\\\nonumber
G^{<0}_{d12\sigma\sigma'\sigma''}(\omega)&=&0, 
G^{>0}_{d12\sigma\sigma'\sigma''}(\omega)=-\delta_{\sigma\sigma'}
\delta_{\sigma\sigma''}2{\pi}i\rho_{21\sigma}\delta(\omega-\epsilon_d-U').\nonumber
\end{eqnarray}
Note that by using the lowest-order gradient expansion with slowly varying in the center-of-mass 
time $T$ and rapidly varying in the relative time $t$ and after the Fourier transformation 
from $t$ to $\omega$, the same results as the above can be acquired.

Inserting Eq.$(13)$ into Eq.$(5)-(10)$, we get the final quantum equations:
\begin{eqnarray}
\dot{\rho}_{00}&=&\sum\limits_{\sigma}[-2\alpha_{1\sigma}\rho_{00}+
beta_{1\sigma}(\rho_{11\sigma}+\rho_{22\sigma})+\beta_{2\sigma}
\rho_{12\sigma}+\beta^{*}_{2\sigma}\rho_{21\sigma}],\\\nonumber
\dot{\rho}_{11\sigma}&=&\alpha_{1\sigma}\rho_{00}-(\beta_{1\sigma}+
\sum\limits_{\sigma'}{\tilde{\alpha}}_{1\sigma'})\rho_{11\sigma}-
1/2(\beta_{2\sigma}+{\tilde{\alpha}}_{2\sigma})\rho_{12\sigma}-
1/2(\beta^{*}_{2\sigma}+{\tilde{\alpha}}^{*}_{2\sigma})\rho_{21\sigma}+
\sum\limits_{\sigma'}{\tilde{\beta}}_{1\sigma'}\rho_{dd\sigma\sigma'},\\\nonumber
\dot{\rho}_{22\sigma}&=&\alpha_{1\sigma}\rho_{00}-(\beta_{1\sigma}+
\sum\limits_{\sigma'}{\tilde{\alpha}}_{1\sigma'})\rho_{22\sigma}-
1/2(\beta_{2\sigma}+{\tilde{\alpha}}_{2\sigma})\rho_{12\sigma}-
1/2(\beta^{*}_{2\sigma}+{\tilde{\alpha}}^{*}_{2\sigma})\rho_{21\sigma}+
\sum\limits_{\sigma'}{\tilde{\beta}}_{1\sigma'}\rho_{dd\sigma'\sigma},\\\nonumber
\dot{\rho}_{21\sigma}&=&\alpha_{2\sigma}\rho_{00}-1/2(\beta_{2\sigma}+
{\tilde{\alpha}}_{2\sigma})(\rho_{11\sigma}+\rho_{22\sigma})-(\beta_{1\sigma}+
\sum\limits_{\sigma'}{\tilde{\alpha}}_{1\sigma'})\rho_{21\sigma}+
{\tilde{\beta}}_{2\sigma}\rho_{dd\sigma\sigma}+i(\epsilon_{1}-\epsilon_{2})
\rho_{21\sigma},\\\nonumber
\dot{\rho}_{dd\sigma\sigma}&=&{\tilde{\alpha}}_{1\sigma}(\rho_{11\sigma}+
\rho_{22\sigma})+{\tilde{\alpha}}_{2\sigma}\rho_{12\sigma}+
{\tilde{\alpha}}^{*}_{2\sigma}\rho_{21\sigma}-2{\tilde{\beta}}_{1\sigma}
\rho_{dd\sigma\sigma},\\\nonumber
\dot{\rho}_{dd\sigma\bar{\sigma}}&=&{\tilde{\alpha}}_{1\sigma}
\rho_{22\bar{\sigma}}+{\tilde{\alpha}}_{1\bar\sigma}\rho_{11\sigma}-
({\tilde{\beta}}_{1\sigma}+{\tilde{\beta}}_{1\bar\sigma})
\rho_{dd\sigma{\bar{\sigma}}}.\nonumber
\end{eqnarray}
Here,
\begin{eqnarray}
\alpha_{1\sigma}&=&f_L(\epsilon_d)\Gamma_{L\sigma}+f_R(\epsilon_d)\Gamma_{R\sigma},
\beta_{1\sigma}=[1-f_L(\epsilon_d)]\Gamma_{L\sigma}+[1-f_R(\epsilon_d)]
\Gamma_{R\sigma},\\\nonumber
{\tilde{\alpha}}_{1\sigma}&=&f_L(\epsilon_d+U')\Gamma_{L\sigma}+
f_R(\epsilon_d+U')\Gamma_{R\sigma},
{\tilde{\beta}}_{1\sigma}=[1-f_L(\epsilon_d+U')]\Gamma_{L\sigma}+
[1-f_R(\epsilon_d+U')]\Gamma_{R\sigma},\\\nonumber
\alpha_{2\sigma}&=&f_L(\epsilon_d)\Gamma_{L\sigma}e^{-i\varphi/2}+
f_R(\epsilon_d)\Gamma_{R\sigma}e^{i\varphi/2},
\beta_{2\sigma}=[1-f_L(\epsilon_d)]\Gamma_{L\sigma}e^{-i\varphi/2}+
[1-f_R(\epsilon_d)]\Gamma_{R\sigma}e^{i\varphi/2},\\\nonumber
{\tilde{\alpha}}_{2\sigma}&=&f_L(\epsilon_d+U')\Gamma_{L\sigma}
e^{-i\varphi/2}+f_R(\epsilon_d+U')\Gamma_{R\sigma}e^{i\varphi/2},\\\nonumber
{\tilde{\beta}}_{2\sigma}&=&[1-f_L(\epsilon_d+U')]\Gamma_{L\sigma}
e^{-i\varphi/2}+[1-f_R(\epsilon_d+U')]\Gamma_{R\sigma}e^{i\varphi/2}.\nonumber
\end{eqnarray}
Complemented with the completeness relation
$\rho_{00}+2\rho_{11}+2\rho_{22}+2\rho_{dd}+2\rho_{d\bar{d}}=1$, the closed equations Eq.$(14)$ can be solved to determine the expectation values of the matrix
in steady states.

\section{Discussion}
In the equilibrium state, only the diagonal distribution probabilities are nonzero,
$\rho_{00}=1/Z, \rho_{11\sigma}=e^{-\beta\epsilon_d}/Z, \rho_{22\sigma}=e^{-\beta\epsilon_d}/Z,
\rho_{dd\sigma\sigma'}=e^{-\beta(2\epsilon_d+U')}/Z$ and $Z=1+4e^{-\beta\epsilon_d}+4e^{-\beta(2\epsilon_d+U')}$,
which can be readily gotten from the above equations set with $f_L(\epsilon_d)=f_R(\epsilon_d)=(e^{\beta\epsilon_{d}}+1)^{-1},
f_L(\epsilon_{d}+U')=f_R(\epsilon_{d}+U')=\{e^{\beta(\epsilon_{d}+U')}+1\}^{-1}$ supposing the equilibrium chemical potential $\mu_L=\mu_R=0$.
The results meet the classical Boltzmann distribution in weak coupling limit.
It is obvious that the occupation number of dot $1$ with certain spin $c^{\dag}_{1\sigma}c_{1\sigma}=\rho_{11\sigma}+\sum_{\sigma'}\rho_{dd\sigma\sigma'}$
has nothing to do with the magnetic flux $\Phi$ and the effect of dot level disparity $\epsilon$ is not taken into account, which is the same as dot $2$, and
in the spin symmetry space $\rho_{dd\sigma\sigma}=\rho_{dd\sigma\bar{\sigma}}$.

Applying external voltage on the two leads, the whole system is driven out of equilibrium and off-diagonal elements play a vital role in transport, whereupon some novel features arise.
In the spin symmetry space $\alpha(\tilde{\alpha})_{1(2)\sigma}$, $\beta(\tilde{\beta})_{1(2)\sigma}$, $\Gamma_{L(R)\sigma}$ are all independent of spin, so for simplicity we subtract
their spin index $\sigma$. After straightforward derivation, the solutions for these closed equations and the final formula of current in steady states are obtained in Appendix.
In the following we will see that the occupation number of the dot $1(2)$ shows explicit AB oscillations originating from the interference of two dot states
$\rho_{12\sigma}$ and $\rho_{21\sigma}$, which is certificated by the existence of phase related coefficients $\alpha_{2\sigma}$, $\beta_{2\sigma}$,
$\tilde{\alpha}_{2\sigma}$, $\tilde{\beta}_{2\sigma}$ in the closed equations only stemming from the existence of $\rho_{12\sigma}$ and $\rho_{21\sigma}$.
Furthermore, $\rho_{dd\sigma\sigma}$ is distinct from $\rho_{dd\sigma\bar{\sigma}}$. This difference can be understood in this way.
Interference between electrons in the leads and in the dots survives only when both dots are empty, or one dot is empty and the other is occupied by or obth are occupied by the same spin as
the electron coming from the lead. So the bias-induced coherence of two dot states $\rho_{12\sigma}$ and $\rho_{21\sigma}$ are only associated with spin diagonal
double-occupied state probability $\rho_{dd\sigma\sigma}$ and not with spin nondiagonal one $\rho_{dd\sigma\bar{\sigma}}$, otherwise the path the incoming electron travels can be
identified. This is distinct from double-dot in series where the path is a definite one including both dots \cite{Dong}.
From the equation of $\rho_{dd\sigma\sigma}$, we can see that effect of $\rho_{12\sigma}$ and $\rho_{21\sigma}$ is to accelerate its change with the time.

\begin{figure}[h]
\includegraphics [width=0.6\textwidth,clip] {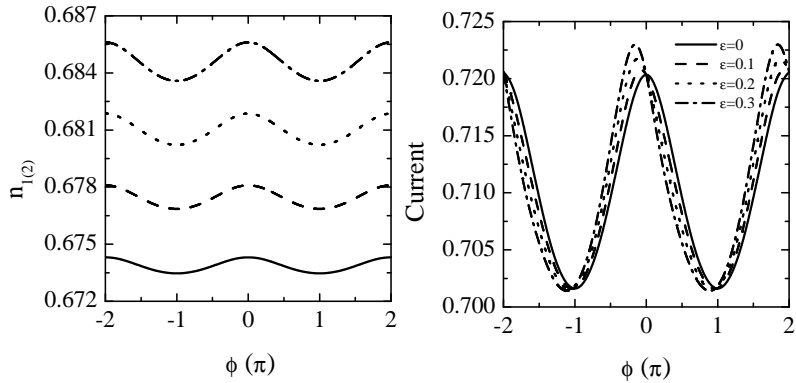}
\caption{Left: AB oscillations of dot population $n_{1(2)}$ with interdot correlation $U'=0$ and energy disparity $\epsilon=0$
at temperature $k_{B}T=20$ and bias voltage $eV=50$ for different average level $\epsilon_d$:
$-1$(solid), $-1.5$(dash), $-2$(dot), $-2.5$(dash dot).
Right: AB oscillations of current for different energy disparity $\epsilon$ with interdot correlation $U'=20$ and average level $\epsilon_d=-1$
at temperature $k_{B}T=20$ and bias voltage $eV=50$. $e=\hbar=\Gamma=1.$}
\label{fig2}
\end{figure}
\vspace{0.2in}
\begin{figure}[h]
\includegraphics [width=0.6\textwidth,clip] {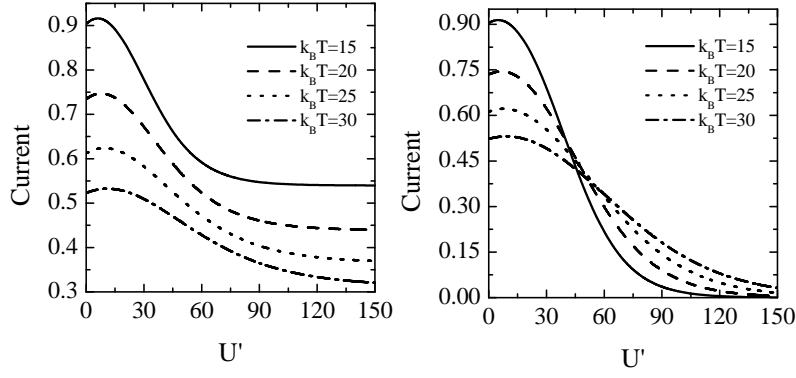}
\caption{Current versus interdot correlation $U'$ with various
temperatures for average dot level $\epsilon_d=-1$,
energy disparity $\epsilon=0$, and bias voltage $eV=50$ at $\varphi=0$
(left) and at $\varphi=\pi$(right). $e=\hbar=\Gamma=1.$} 
\label{fig3}
\end{figure}

Note that the temperature is assumed higher than the Kondo temperature so that the Kondo correlation which
is attributed to the strong coupling between leads and dots can be negligible.
In the sequential tunneling picture, the tunneling contribution to transport is up
to first order in $\Gamma$ in the Coulomb blockade regime.
We can directly calculate the current from Eq.$(21)$ which is valid for arbitrary bias voltage $eV$, interdot repulsion $U'$,
and dot energy position $\epsilon_d$ in broad ranges and for tiny level variation $\epsilon$ between two dots.
For convenience, the symmetric bias voltage $\mu_L=-\mu_R=eV/2$ is applied, then all the results
are invariant under the reversals of magnetic flux $\varphi{\rightarrow}-\varphi$ and of bias voltage $eV{\rightarrow}-eV$ simultaneously.
This is in accordance with the Onsager relation for two-terminal setups \cite{JKO}.
In general, the occupation number of dot $1(2)$ and the current through the system all have AB oscillations, combinations of $\cos{\varphi}$ and $\sin{\varphi}$,
with a period of $\Phi_0$, and the amplitudes depend on the resonant energy level
$\epsilon_d$ and $\epsilon_d+U'$ relative to the chemical potentials of two leads and small disparity amplitude $\epsilon$.
It is also supposed that $k_{B}T{\gg}\Gamma$, $|\epsilon_1|$, $|\epsilon_2|$, and $\Gamma{\gg}|\epsilon|$,
so the lowest-order transport dominates.
In Fig.$2$ we plot AB oscillations of the population in dot with different level positions and the current with different energy
discrepancys respectively. For two identical dots $\epsilon=0$, all oscillations are symmetric about $\varphi=0$
including only $\cos{\varphi}$ with larger amplitudes for deeper levels. The population reaches maximum at $\varphi=0$ and
minimum at $\varphi=\pi$. For two distinct dots $\epsilon{\not=}0$, the symmetry
about $\varphi=0$ is lost, which gets more apparent with larger $\epsilon$ for oscillations following $\sin{\varphi}$ take effect,
and the extremes at $\varphi=0$ and $\varphi=\pi$ also disappear.
Fig.$3$ shows current versus interdot correlation with various temperatures for $\varphi=0$ and $\varphi=\pi$.
We can see that at finite temperature and finite bias the interdot correlation does not always inhibit
transport. Only when $U'$ is large enough, the prohibition effect appears, especially at
$\varphi=\pi$, and this is mitigated by raising the temperature.

In the following two specially simple cases with equal energy levels of two dots are under consideration.
First is that doubly occupied states are forbidden by infinite interdot Coulomb repulsion $U'$, such that
$\rho_{dd\sigma\sigma'}=0$, and $\tilde{\alpha}_1=0, \tilde{\alpha}_2=0$.
The steady solutions are
\begin{eqnarray}
\rho_{11}&=&\frac{(2-f_L-f_R)[(1-f_L)f_R+(1-f_R)f_L]}{2(f_L+f_R-3f_Lf_R+1)
(2-f_L-f_R)},\\\nonumber
\rho_{12}&=&\frac{f_L-f_R}{2(f_L+f_R-3f_{L}f_{R}+1)}[\frac{f_R-f_L}{2-f_L-f_R}\cos{\frac{\varphi}{2}}
+i\sin{\frac{\varphi}{2}}],\\\nonumber
\rho_{00}&=&\frac{f_{L}f_{R}-f_L-f_R+1}{f_L+f_R-3f_{L}f_{R}+1},\\\nonumber
\end{eqnarray}
and the current is
\begin{eqnarray}
I=\frac{2{\pi}e\Gamma}{h}\frac{2(f_L-f_R)(1-f_R)(1-f_L)}{(f_L+f_R-3f_{L}f_{R}+1)
(2-f_L-f_R)}(1+\cos{\varphi}).
\end{eqnarray}
It is evident that in this situation transport through the whole system is totally coherent,
even though the population of dot $1(2)$ shows no AB oscillations.
With $\varphi=0, {\pm}2\pi, {\pm}4\pi,...$, the current reaches the maximum, and phase locking always happens
for any voltage and dot levels. With $\varphi={\pm}\pi, {\pm}3\pi, {\pm}5\pi,...$, the current vanishes and the
interference between two dots is completely destructive, since the whole system now is equivalent to the sum of two separated subsystems
and no transport will take place for any bias, which conforms to Ref.\cite{JKO}.
Considering only the first order of $\Gamma$ in sequential tunneling regime, we get the full coherence.

Second is that empty state is forbidden in deep level status with $\epsilon_d$ far below the Fermi level and $\epsilon_{d}+U'$ just above it in equilibrium state,
so that $\rho_{00}=0$, and $\beta_1=0, \beta_2=0$.
The steady solutions are
\begin{eqnarray}
\rho_{11}&=&\frac{(2-\tilde{f}_L-\tilde{f}_R)[(1-\tilde{f}_L)(2\tilde{f}_R+
\tilde{f}_L-\tilde{f}_R\cos{\varphi})+(1-\tilde{f}_R)(2\tilde{f}_L+\tilde{f}_R-
\tilde{f}_L\cos{\varphi})]}
{4(\tilde{f}_L+\tilde{f}_R)(3-\cos{\varphi})(2-\tilde{f}_L-\tilde{f}_R)+
2(\tilde{f}_L-\tilde{f}_R)^2(1-\cos{\varphi})},\\\nonumber
\rho_{12}&=&\frac{1}{4(\tilde{f}_L+\tilde{f}_R)}[(2-\tilde{f}_L-
\tilde{f}_R)\cos{\frac{\varphi}{2}}+i\sin{\frac{\varphi}{2}}(\tilde{f}_L-\tilde{f}_R)-
(4\cos{\frac{\varphi}{2}}-2i\sin{\frac{\varphi}{2}}\frac{\tilde{f}_R-\tilde{f}_L}{2-
\tilde{f}_L-\tilde{f}_R})\\\nonumber
&&\frac{(2-\tilde{f}_L-\tilde{f}_R)[(1-\tilde{f}_L)(2\tilde{f}_R+
\tilde{f}_L-\tilde{f}_R\cos{\varphi})+(1-\tilde{f}_R)(2\tilde{f}_L+
\tilde{f}_R-\tilde{f}_L\cos{\varphi})]}
{4(\tilde{f}_L+\tilde{f}_R)(3-\cos{\varphi})(2-\tilde{f}_L-\tilde{f}_R)+
2(\tilde{f}_L-\tilde{f}_R)^2(1-\cos{\varphi})}],\\\nonumber
\rho_{dd}&=&\frac{1}{2}-\frac{(4-\tilde{f}_L-\tilde{f}_R)[(1-\tilde{f}_L)
2\tilde{f}_R+\tilde{f}_L-\tilde{f}_R\cos{\varphi})+(1-\tilde{f}_R)(2\tilde{f}_L+
\tilde{f}_R-\tilde{f}_L\cos{\varphi})]}
{4(\tilde{f}_L+\tilde{f}_R)(3-\cos{\varphi})(2-\tilde{f}_L-\tilde{f}_R)+
2(\tilde{f}_L-\tilde{f}_R)^2(1-\cos{\varphi})},\\\nonumber
\rho_{d\bar{d}}&=&\frac{(\tilde{f}_L+\tilde{f}_R)[(1-\tilde{f}_L)(2\tilde{f}_R+
\tilde{f}_L-\tilde{f}_R\cos{\varphi})+(1-\tilde{f}_R)(2\tilde{f}_L+\tilde{f}_R-
\tilde{f}_L\cos{\varphi})]}
{4(\tilde{f}_L+\tilde{f}_R)(3-\cos{\varphi})(2-\tilde{f}_L-\tilde{f}_R)+
2(\tilde{f}_L-\tilde{f}_R)^2(1-\cos{\varphi})},\\\nonumber
\end{eqnarray}
and the current is
\begin{eqnarray}
I&=&\frac{{\pi}e\Gamma}{h}\frac{\tilde{f}_L-\tilde{f}_R}{\tilde{f}_L+\tilde{f}_R}
\{1+2(\tilde{f}_L+\tilde{f}_R)+(1-\tilde{f}_L-\tilde{f}_R)\cos{\varphi}
-[(4-\tilde{f}_L-\tilde{f}_R)\\\nonumber&+&(4-3\tilde{f}_L-3\tilde{f}_R)\cos{\varphi}]
\frac{(1-\tilde{f}_L)(2\tilde{f}_R+\tilde{f}_L-\tilde{f}_{R}\cos{\varphi})+
(1-\tilde{f}_R)(2\tilde{f}_L+\tilde{f}_R-\tilde{f}_{L}\cos{\varphi})}
{2(\tilde{f}_L+\tilde{f}_R)(3-\cos{\varphi})(2-\tilde{f}_L-\tilde{f}_R)+
(\tilde{f}_L-\tilde{f}_R)^{2}(1-\cos{\varphi})} \},
\end{eqnarray}
which is partially coherent, and the AB oscillations for the population in dots and the current all follow the $\cos{\varphi}$ form.

\section{Summary}
To conclude, we have studied the quantum transport through a parallel double-dot structure with the Aharonov-Bohm magnetic flux infiltering the
two path closed region. Infinite intradot and arbitrary interdot Coulomb repulsion have been considered by employing the slave-boson
technique introduced by Zou and Anderson. In weak coupling and sequential tunneling regime we use the "classical" quantum rate equations combined
with nonequilibrium Green's functions to determine the density matrix and to calculate the current flowing through the system of stationary state.
The results show that external bias voltage induced superposition of two dots states is the cause of phase coherence of AB oscillations of population
of each dot and the current as a function of magnetic flux having a period of $\Phi_0$. Two simplest cases are discussed as
examples for two identical dots. We find that if there are no doubly occupied states, the current is totally coherent - completely destructive interference exists with $\varphi$
odd multiples of $\pi$ and maximum always reaches with $\varphi$ even multiples of $\pi$; if there is no empty state, the population and the current are partially coherent
and amplitudes depend on the resonant energy levels relative to the chemical potentials of two reservoirs.

This work was supported by the National Science Foundation of China,
the Special Funds for Major State Basic Research Project, and
the Shanghai Municipal Commission of Science and Technology.

\section{Appendix}
The solutions of density matrix elements in stationary states are: 
$\rho_{11\sigma}=\rho_{11\bar{\sigma}}=\rho_{11}$, $\rho_{22\sigma}=\rho_{22\bar{\sigma}}=\rho_{22}$,
$\rho_{21\sigma}=\rho_{21\bar{\sigma}}=\rho_{21}$, $\rho_{12\sigma}=\rho_{12\bar{\sigma}}=\rho_{12}$,
$\rho_{dd\sigma\sigma}=\rho_{dd\bar{\sigma}\bar{\sigma}}=\rho_{dd}$,
$\rho_{dd\sigma\bar{\sigma}}=\rho_{dd\bar{\sigma}\sigma}=\rho_{d\bar{d}}$,
\begin{eqnarray}
\rho_{00}&=&{(C_{2}A_{1}-C_{1}A_{2})}/{(B_{1}A_{2}-B_{2}A_{1})},\\\nonumber
\rho_{11}&=&{(C_{2}B_{1}-C_{1}B_{2})}/{[2(A_{1}B_{2}-A_{2}B_{1})]},\\\nonumber
\rho_{21}&=&\{{(2\alpha_2-\tilde{\beta}_2)
\rho_{00}-[\tilde{\alpha}_2+\beta_2+(2+{{\tilde{\alpha}}_1}/{{\tilde{\beta}}_1})
\tilde{\beta}_2](\rho_{11}+\rho_{22})+{\tilde{\beta}}_2}\}/{[2(\beta_1+
2{\tilde{\alpha}}_{1}+i{\epsilon})]},\\\nonumber
\rho_{dd}&=&{[1-\rho_{00}-(2+{{\tilde{\alpha}}_1}/{{\tilde{\beta}}_1})(\rho_{11}+
\rho_{22})]}/{2},\\\nonumber
\rho_{d\bar{d}}&=&{{\tilde{\alpha}}_1(\rho_{11}+\rho_{22})}/{(2{\tilde{\beta}}_1)},\nonumber
\end{eqnarray}
and $\rho_{22}=\rho_{11}$, $\rho_{12}=\rho^{*}_{21}$.
Here,
\begin{eqnarray}
A_{1}&=&4({\Delta}^2+{\epsilon}^2)-\Delta(\tilde{f}_L+\tilde{f}_R\cos{\varphi})
[1+\tilde{f}_L-f_L+\gamma(1-\tilde{f}_L)]\nonumber
-\Delta(\tilde{f}_R+\tilde{f}_L\cos{\varphi})[1+\tilde{f}_R-f_R
\\\nonumber&+&\gamma(1-\tilde{f}_R)]
+{\epsilon}\sin{\varphi}[(\tilde{f}_R-\tilde{f}_L)(1+\gamma)+\tilde{f}_{L}f_R-
\tilde{f}_{R}f_L],\\\nonumber
B_{1}&=&\Delta[2\tilde{f}_L(f_L+f_R\cos{\varphi})+2\tilde{f}_R(f_R+f_L\cos{\varphi})-
(1-\tilde{f}_L)(\tilde{f}_L+\tilde{f}_R\cos{\varphi})-(1-\tilde{f}_R)(\tilde{f}_R
\\\nonumber&+&\tilde{f}_L\cos{\varphi})]
+(2-\tilde{f}_L-\tilde{f}_R)({\Delta}^2+{\epsilon}^2)-\epsilon\sin{\varphi}
[2(\tilde{f}_{R}f_L-\tilde{f}_{L}f_R)+\tilde{f}_L-\tilde{f}_R],\\\nonumber
C_{1}&=&-(2-\tilde{f}_L-\tilde{f}_R)({\Delta}^2+{\epsilon}^2)+
\Delta[(1-\tilde{f}_L)(\tilde{f}_L+\tilde{f}_R\cos{\varphi})+
(1-\tilde{f}_R)(\tilde{f}_R+\tilde{f}_L\cos{\varphi})]\\\nonumber
&+&\epsilon\sin{\varphi}(\tilde{f}_L-\tilde{f}_R),\\\nonumber
A_{2}&=&(2-f_L-f_R)({\Delta}^2+{\epsilon}^2)
-\Delta(1-f_L)[1+\tilde{f}_L-f_L+\cos{\varphi}(1+\tilde{f}_R-f_R)+
\gamma(1-\tilde{f}_L+\cos{\varphi}\\\nonumber&-&\cos{\varphi}\tilde{f}_R)]
-\Delta(1-f_R)[1+\tilde{f}_R-f_R+\cos{\varphi}(1+\tilde{f}_L-f_L)+
\gamma(1-\tilde{f}_R+\cos{\varphi}-\cos{\varphi}\tilde{f}_L)]\\\nonumber
&+&\epsilon\sin{\varphi}[(1-f_R)(\tilde{f}_L+\gamma-
\gamma\tilde{f}_L)-(1-f_L)(\tilde{f}_R+\gamma-\gamma\tilde{f}_R)],\\\nonumber
B_{2}&=&\Delta(2f_L+\tilde{f}_L-1)(1-f_L+\cos{\varphi}-\cos{\varphi}f_R)+
\Delta(2f_R+\tilde{f}_R-1)(1-f_R+\cos{\varphi}-\cos{\varphi}f_L)\\\nonumber
&-&2(f_L+f_R)({\Delta}^2+{\epsilon}^2)-\epsilon\sin{\varphi}
[(1-f_R)(1+\tilde{f}_L)-(1-f_L)(1+\tilde{f}_R)],\\\nonumber
C_{2}&=&\Delta[(1-f_L)(1-\tilde{f}_L+\cos{\varphi}-\cos{\varphi}\tilde{f}_R)+
(1-f_R)(1-\tilde{f}_R+\cos{\varphi}-\cos{\varphi}\tilde{f}_L)]\\\nonumber
&-&\epsilon\sin{\varphi}[(1-f_R)(1-\tilde{f}_L)-(1-f_L)(1-\tilde{f}_R)],\nonumber
\end{eqnarray}
and the current formula becomes
\begin{eqnarray}
I&=&\frac{2{\pi}e\Gamma}{h}[(f_L-f_R)(2\rho_{00}+\rho_{11}+\rho_{22})
+\cos{\varphi/2}(f_L+\tilde{f_L}-f_R-\tilde{f_R})(\rho_{12}+\rho_{21})\\\nonumber
&-&i\sin{\varphi/2}
(f_L+\tilde{f_L}+f_R+\tilde{f_R}-2)(\rho_{12}-\rho_{21})+2(\tilde{f_L}-\tilde{f_R})
(\rho_{11}+\rho_{22}+\rho_{dd}+\rho_{d\bar{d}})],
\end{eqnarray}
where $\Delta{\equiv}2-f_L-f_R+2\tilde{f}_L+2\tilde{f}_R,\  \gamma{\equiv}(4-\tilde{f}_L-\tilde{f}_R)/(2-\tilde{f}_L-\tilde{f}_R),$\\
and $f_{L(R)}{\equiv}f_{L(R)}(\epsilon_d)=\{e^{\beta(\epsilon_{d}-\mu_{L(R)})}+1\}^{-1},\ \tilde{f}_{L(R)}{\equiv}f_{L(R)}(\epsilon_d+U')=\{e^{\beta(\epsilon_{d}+U'-\mu_{L(R)})}+1\}^{-1}$.

\end{document}